\documentclass[fleqn,10pt]{wlscirep}
\usepackage[utf8]{inputenc}
\usepackage[T1]{fontenc}
\usepackage{multirow}

\usepackage{xurl}

\title{CheXmask: a large-scale dataset of anatomical segmentation masks for multi-center chest x-ray images}

\author[1]{Nicol\'as Gaggion}
\author[2, 3]{Candelaria Mosquera}
\author[1]{Lucas Mansilla}
\author[4]{Julia Mariel Saidman}
\author[4]{Martina Aineseder}
\author[1]{Diego H. Milone}
\author[1,*]{Enzo Ferrante}

\affil[1]{Institute for Signals, Systems and Computational Intelligence, sinc(i) CONICET-UNL, Santa Fe, CP 3002, Argentina}
\affil[2]{Health Informatics Department at Hospital Italiano de Buenos Aires, Buenos Aires, CP 1199, Argentina}
\affil[3]{Universidad Tecnológica Nacional, Buenos Aires, CP 1179, Argentina}
\affil[4]{Radiology Department, Hospital Italiano de Buenos Aires, Buenos Aires, CP 1199, Argentina}

\affil[*]{corresponding author: Enzo Ferrante (eferrante@sinc.unl.edu.ar)}

\begin{abstract}
The development of successful artificial intelligence models for chest X-ray analysis relies on large, diverse datasets with high-quality annotations. While several databases of chest X-ray images have been released, most include disease diagnosis labels but lack detailed pixel-level anatomical segmentation labels. To address this gap, we introduce an extensive chest X-ray multi-center segmentation dataset with uniform and fine-grain anatomical annotations for images coming from \textcolor{black}{five} well-known publicly available databases: ChestX-ray8, CheXpert, MIMIC-CXR-JPG, Padchest, and VinDr-CXR, resulting in 657,566 segmentation masks. Our methodology utilizes the HybridGNet model to ensure consistent and high-quality segmentations across all datasets. Rigorous validation, including expert physician evaluation and automatic quality control, was conducted to validate the resulting masks. Additionally, we provide individualized quality indices per mask and an overall quality estimation per dataset. This dataset serves as a valuable resource for the broader scientific community, streamlining the development and assessment of innovative methodologies in chest X-ray analysis. 
The CheXmask dataset is publicly available at: \url{https://physionet.org/content/chexmask-cxr-segmentation-data/}.

\end{abstract}
\begin{document}

\flushbottom
\maketitle

\thispagestyle{empty}


\section*{Background \& Summary}

Chest radiography is a pivotal imaging technique used to diagnose a variety of lung diseases, including pneumonia, tuberculosis, and lung cancer. The significant role of chest X-rays (CXR) in clinical practice is ascribed to their non-invasive nature, relatively low cost, and rapid diagnostic potential. However, the interpretation of these images poses a considerable challenge due to the intricate and overlapping structures within the thoracic cavity, and the subtle manifestations of certain pathological conditions. The high demand for chest radiography and the global shortage of radiologists accentuate the need for efficient and reliable automated analysis systems.

In recent years, methods based on deep learning (DL) have demonstrated exceptional prowess in interpreting medical images, rivaling and occasionally surpassing expert human performance \cite{rajpurkar2017chexnet, irvin2019chexpert}. Convolutional neural networks (CNN) have been particularly instrumental in facilitating such computer-aided diagnosis (CADx) systems \cite{shen2017deep, litjens2017survey}. Nonetheless, the success of these algorithms is closely tethered to the availability of accurately annotated data, with sufficient quantity and diversity, to train the models.

An essential task within this framework is segmentation – the delineation of specific anatomical structures or pathological lesions within an image. In the context of CXR, this might involve the demarcation of anatomical structures such as lungs or heart, or the location of disease abnormalities \cite{roulet2019joint}. Accurate and robust segmentation can serve as a precursor to other downstream tasks, for example providing significant information about the location and size of specific organs or detected abnormalities. However, manual segmentation is a time-consuming process, demanding substantial expertise, and thus, does not scale well to the size of large databases required for DL model training \cite{ronneberger2015u}. 

HybridGNet, a deep learning model for realistic organ contouring, offers a solution for the generation of anatomically plausible CXR segmentations \cite{gaggion2021,gaggion2022}. Utilizing a hybrid approach, it combines conventional convolution operations for image encoding with graph generative models for the anatomically-guided delineation of organ contours. The HybridGNet model was initially introduced with a small CXR landmark dataset to demonstrate its efficacy. In this work, we leverage this model to accomplish our main objective: introducing a large-scale segmentation dataset, named CheXmask, which provides anatomical masks with their corresponding quality index, for 5 extensive chest X-ray databases: Chest x-ray8\cite{wang2017chestxray8}, Chexpert\cite{irvin2019chexpert}, MIMIC-CXR-JPG\cite{johnson2019mimic-jpg}, Padchest\cite{bustos2020padchest} and VinDr-CXR\cite{nguyen2022vindr}. These databases collectively represent a wide variety of geographical locations, patient demographics, and disease spectra, enabling the development of a broad, diverse segmentation dataset. 

As the original databases lack manually curated ground-truth segmentations, we perform quality control by implementing our own Reverse Classification Accuracy (RCA) framework \cite{valindria2017rca}. RCA allows to estimate the accuracy of a segmentation method for an individual image with no ground-truth (GT) masks, which is particularly valuable for large-scale image analysis studies like ours. The fundamental concept behind RCA involves training an auxiliary model (known as the reverse classifier) solely on the individual image, using its predicted segmentation as pseudo-GT. This model is then evaluated on a reference database that contains GT data to obtain a performance metric, which is expected to correlate with the performance that would be measured for the individual image if its GT was available. We validated this method by comparing it to traditional performance evaluation on a subset of test images with masks manually segmented by an expert physician. Additionally, since large-public CXR databases built from automatic analysis of electronic health records (EHR) are subject to errors both in image selection and image annotation, we found that RCA is a useful tool to detect out-of-distribution samples (e.g. poor-quality images). Thus, the RCA metrics for HybridGNet segmentations stand out as a powerful quality metric to handle large databases for downstream tasks, by detecting not only low quality segmentation masks, but also images that should be filtered out.

Our comprehensive analysis underscores the capacity of the HybridGNet model to generate high-quality segmentations of lungs and heart structures in CXR, and presents the RCA method as a way to use these segmentations for detection of poor-quality images in large volumes of data. CheXmask dataset provides a key resource to the medical imaging community and represents a significant stride towards democratizing access to diverse, large-scale segmentation datasets, thereby propelling the advancement of automated CXR analysis research.

\textcolor{black}{In this context, it is important to notice that there are limited publicly available chest X-ray datasets with expert segmentation masks, and most studies mainly rely on the Shenzhen Hospital \cite{shenzhen}, Montgomery County \cite{montgomeryset}, and JSRT \cite{jsrt_shiraishi2000development} datasets. Taken together, they correspond to less than 1000 annotations and the segmented structures are not consistent (for example, while JSRT includes lung, hearth and clavicle annotations, Shenzhen and Montgomery only include lung annotations). Concurrently to our work, larger datasets have recently emerged, like the one released by Seibold and co-workers \cite{seibold2023accurate}. However, even though this dataset provides segmentations for a large number of X-ray images featuring different anatomical structures, they are not real X-ray images, since they were generated as digitally reconstructed radiographs, by projecting CT images from the PAX-Ray++ Dataset \cite{Seibold_2022_BMVC}. As highlighted by the authors, real X-ray images tend to use a harder type of radiation than standard CTs. This fact, together with the contrast agents used in CTs, might lead to a divergence of visual quality between the projected X-rays and real ones \cite{seibold2023accurate}. Moreover, they only provide pixel-level masks, while we also provide landmark based segmentations with correspondences. In this context, CheXmask stands out as a unique large scale dataset for real X-ray images, with quality control validated through objective indices.}

\section*{Methods}



\begin{figure}
    \centering
    \includegraphics[width=0.75\linewidth]{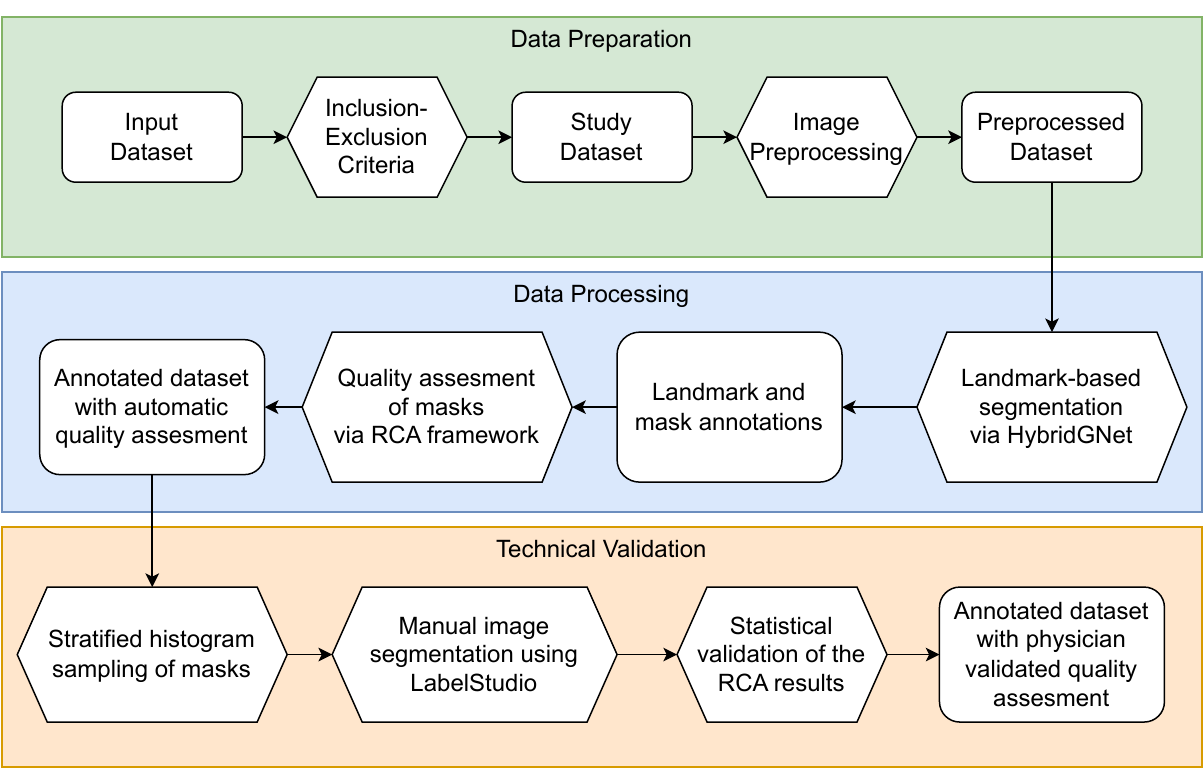}
    \caption{Data processing flowchart depicting the main steps involved in the building of the CheXmask dataset.}
    \label{fig:flowchart}
\end{figure}

\subsection*{Data Preparation}

\subsubsection*{Image datasets}

In this study, we utilized five extensive CXR datasets: ChestX-ray8 (available in a dedicated website \cite{chestxray8_dataset}), CheXpert (available in a dedicated website \cite{chexpert_dataset}), MIMIC-CXR-JPG (available at PhysioNet\cite{mimiccxrjpg_dataset}), Padchest (available in a dedicated website \cite{padchest_dataset}), and VinDr-CXR (available at PhysioNet\cite{vindrcxr_dataset}), \textcolor{black}{original sources to download each dataset are provided in the web references}. Figure \ref{fig:flowchart} provides an overview of the complete study, which was repeated five times, one for each dataset individually. The ChestX-ray8 dataset \cite{wang2017chestxray8} contains 112,120 frontal-view X-ray images from 30,805 unique patients, annotated with text-mined fourteen disease image labels. The CheXpert dataset \cite{irvin2019chexpert} comprises 224,316 chest radiographs collected from Stanford Hospital between October 2002 and July 2017. The MIMIC-CXR-JPG dataset \cite{johnson2019mimic-jpg} includes 377,110 CXR images from 227,827 imaging studies involving 65,379 patients; these were collected at the Beth Israel Deaconess Medical Center Emergency Department in the United States from 2011 to 2016. The Padchest dataset \cite{bustos2020padchest} is composed of 160,861 images from over 67,000 patients, collected from Hospital San Juan in Spain from 2009 to 2017. Lastly, the VinDr-CXR dataset \cite{nguyen2022vindr} consists of 18,000 manually annotated images gathered from two primary hospitals in Vietnam.

\subsubsection*{Inclusion-exclusion criteria}

\begin{figure}
    \centering
    \includegraphics[width=\linewidth]{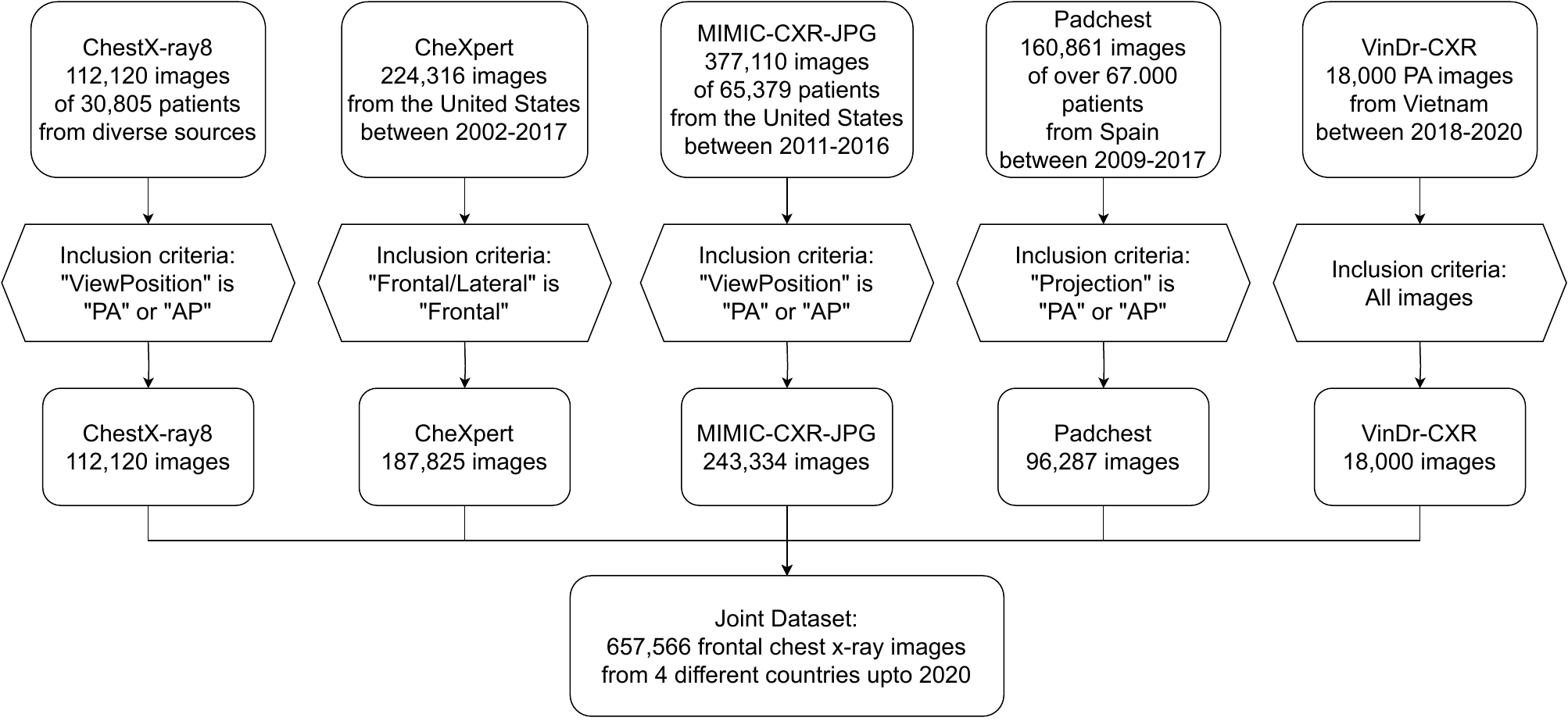}
    \caption{Summary diagram of inclusion-exclusion criteria.}
    \label{fig:inclusion-exclusion}
\end{figure}

Our study incorporated only frontal images, captured in posteroanterior (PA) or anteroposterior (AP) views (no lateral views were included). Specific selection criteria varied among the datasets due to their differing metadata, as summarized in Figure \ref{fig:inclusion-exclusion}.

ChestX-ray8, and VinDr-CXR datasets contain only frontal images, so all images were included. We incorporated images from the Padchest dataset based on the ``Projection'' metadata field, including only those labeled as PA or AP, reducing the initial count of 160,861 images to 96,287. For the CheXpert dataset, images labeled as ``Frontal'' in the ``Frontal/Lateral'' metadata column were included, resulting in 187,825 out of the total 224,316 images. Lastly, images from the MIMIC-CXR-JPG dataset were included if the ``ViewPosition'' tag was listed as ``PA'' or ``AP'', which resulted in 243.334 out of the initial 377,110 images.

\subsubsection*{Image preprocessing}

For compatibility with HybridGNet input format, we preprocessed images to attain a uniform size suitable for generating pseudo-landmark contours. The standard size of the training data was 1024x1024 pixels, obtained by padding the images to square dimensions then resizing them to the required size, when needed.

Images from ChestX-ray8 dataset already satisfied the required dimension and format, thus requiring no pre-processing. CheXpert, MIMIC-CXR-JPG, Padchest, and VinDr-CXR datasets underwent a standard pre-processing pipeline, which included padding the images to a square shape and then resizing them to 1024x1024 pixels. Additionally, VinDr-CXR images needed extraction from DICOM files before the pre-processing. All these pre-processing procedures are reversible, which allows for the preservation of the original image shapes for releasing the generated annotations.

\subsection*{Data processing}

\subsubsection*{Anatomically plausible segmentation via HybridGNet}

The DL model HybridGNet segments organ contours by detecting the coordinates of anatomical landmarks. Thus, it is trained to minimize the distance between the predicted landmark positions and their GT location. The model incorporates an encoder-decoder architecture that combines standard convolutions for image encoding and graph generative models to achieve anatomically plausible representations. Pixel-level masks are obtained by filling in the contours defined by the landmarks predicted by HybridGNet. A detailed description of the HybridGNet model can be found in the work of Gaggion et al. \cite{gaggion2021,gaggion2022}. A modified training procedure was presented in a later work \cite{gaggionISBI2023}, to avoid domain memorization issues which emerge when dealing with heterogeneous labels in multi-centric scenarios (i.e. images from some medical centers contain only lung annotations, while others include both lung and heart masks).

In this work, we used the modified multi-centric training procedure \cite{gaggionISBI2023} to retrain the HybridGNet model with the complete dataset used in prior studies (i.e., both the train and test splits of previous works). This dataset, which will be referred to as Chest-Xray-Landmark dataset, was released previously and is available on GitHub: \url{https://github.com/ngaggion/Chest-xray-landmark-dataset}. The Chest-Xray-Landmark dataset is curated from a variety of sources, providing annotations for a total of 911 images. These images are sourced from different repositories, including the JSRT dataset \cite{jsrt_shiraishi2000development} (available from its dedicated website \cite{jsrt_dataset}), a small subset of the Padchest dataset \cite{bustos2020padchest} (accessible on its designated website \cite{padchest_dataset}), as well as the Montgomery \cite{montgomeryset} and Shenzhen \cite{shenzhen} datasets (both obtainable from the NIH National Library of Medicine \cite{montgomery_and_shenzhen_datasets}). Please note that the Chest-Xray-Landmark dataset does not provide the images themselves; users are required to source the original images from their respective origins. All 911 images within this dataset contain lung annotations and only 383 images also include heart labels. The training process incorporated various data augmentation techniques, such as random rotations, scaling, and color shifting, to enhance robustness and generalizability.

A fully working demo of the model used to segment the complete dataset is available on Hugging Face Spaces at \url{https://huggingface.co/spaces/ngaggion/Chest-x-ray-HybridGNet-Segmentation}.

\subsubsection*{Reverse classification accuracy to measure segmentation quality}

We employed RCA to estimate the Dice Similarity Coefficient (DSC) \cite{dice1945measures} in the absense of ground truth. DSC is a widely used metric for assessing the similarity between two sets of binary labels. By using RCA, it is possible to produce accurate estimations of DSC, thereby making it easier to perform quality control in the absence of GT. The DSC quantifies the overlap between the predicted segmentation and the GT segmentation, ranging from 0 (no overlap) to 1 (perfect overlap). It is defined as $ DSC = \frac{2TP}{2TP+FP+FN}$, where $TP$, $FP$, and $FN$ denote the number of true positive, false positive, and false negative pixels, respectively. 

To obtain the estimated DSC for a segmentation mask, the RCA framework consists in training a new segmentation method using solely this image with the predicted mask as GT. This new trained model (i.e., the reverse classifier) is then applied to segment a reference image set, with known GT masks. The hypothesis is that the segmentation accuracy of the reverse model on the reference set correlates to the segmentation quality of the original image: if the original segmentation of the evaluated image was good, the segmentation accuracy of the reverse model on this reference set should also be good. 

\begin{figure}
    \centering
    \includegraphics[width=0.8\linewidth]{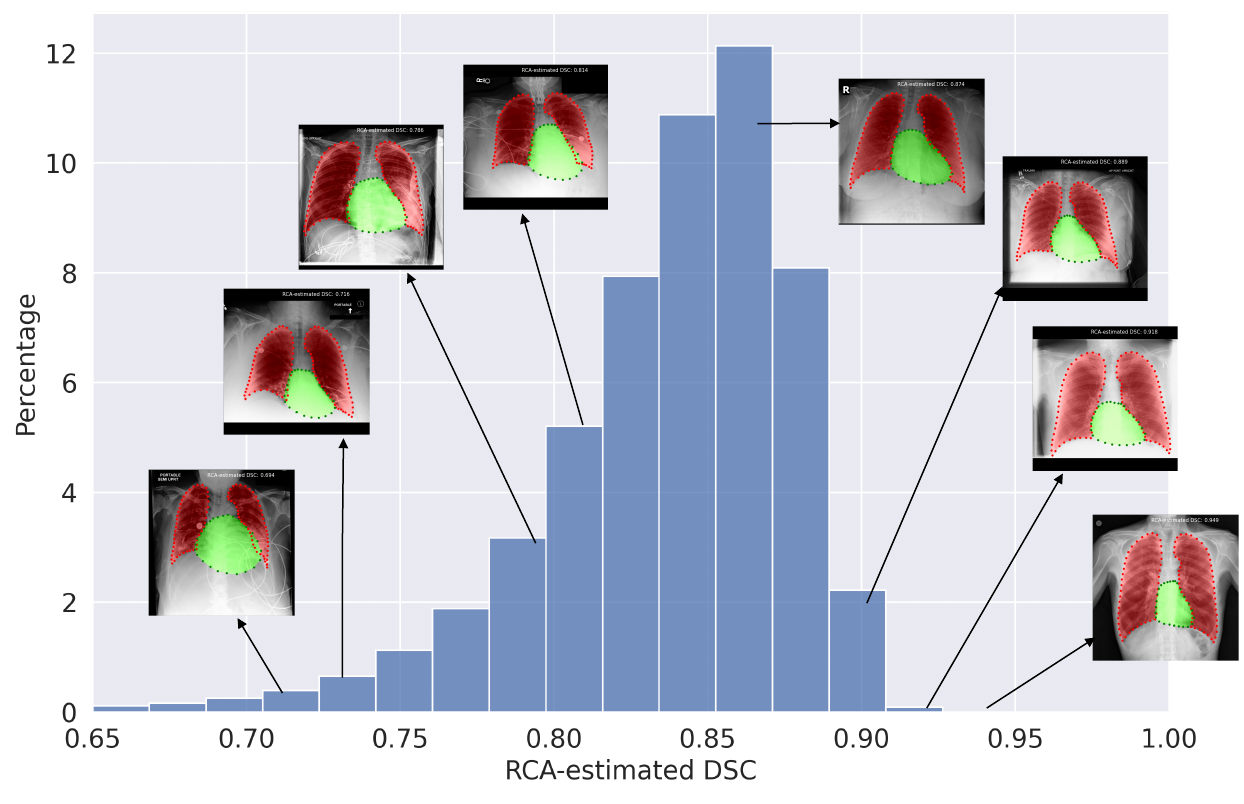}
    \caption{Histogram showcasing the distribution of the RCA-estimated DSC for the complete CheXmask database. Example images with their landmark-based segmentations were drawn with lines to their corresponding histogram bin. \textcolor{black}{Reproduction of x-ray images was allowed upon request to the original sources.}}
    \label{fig:histogram_complete}
\end{figure}

In the original work, the authors experiment with three different methods for implementing the RCA classifier: Atlas Forests, Deep Learning, and Atlas-based Label Propagation. In this case, we used the last one, which employs a non-rigid registration technique for single-atlas label propagation, where the single atlas corresponds to the evaluated image together with its predicted segmentation. A summary value of the distribution of DSC across the reference set images is then used as performance metric. We will use for these values the \emph{mean} and \emph{max} RCA-estimated DSC. The RCA-estimated DSC is expected to correlate well with the real DSC that one would obtain if GT data was available. Figure \ref{fig:histogram_complete} showcases the distribution of RCA-estimated DSC for the complete CheXmask dataset with examples sampled from each bin.

The use of DL in atlas registration has shown improved performance over classic multi-atlas approaches, while also offering a significant speed-up in computation. In order to speed up the evaluation process, we implemented a DL-based atlas registration procedure \cite{mansilla2020learning} as label propagation method (that we refer as Deep RCA), which is several orders of magnitude faster than traditional registration methods. Specifically, we divide the registration procedure into two stages: first performing rigid registration to globally align the corresponding images, and then proceeding with deformable registration to correct local misalignments.

\section*{Data Records}

The CheXmask dataset is structured as one file of comma-separated values (CSV) for each CXR dataset. The CSV content is explained in Table \ref{table:data_records_1}. We do not include images or metadata from the original datasets. We include an image ID in the first column of the CSV files, which has the same column name as the ID column of the respective dataset, and allows to match the rows in this CSV with the original dataset's instances. We also include the pre-processed version of the masks, allowing to use all datasets in the same image resolutions (please also note that ChestX-ray8 masks are already in the desired resolution). This database \cite{gaggion2023chexmask} is available at PhysioNet \cite{goldberger2000physiobank}, in the following repository: \url{https://physionet.org/content/chexmask-cxr-segmentation-data/}. 

\begin{table}[h!]
\centering
\caption{Description of the columns in CheXmask CSV files.}
\begin{tabular}{l p{12cm}}
\hline
\textbf{Column} & \textbf{Description} \\
\hline
Image ID & Contains references to the original images as per the original metadata, thus the column name changes across dataset. \\
Dice RCA (Max) & Provides the maximum Dice Similarity Coefficient for the Reverse Classification Accuracy (RCA), indicating the quality of the segmentation. \\
Dice RCA (Mean) & Provides the mean Dice Similarity Coefficient for the Reverse Classification Accuracy (RCA), indicating the quality of the segmentation. \\
Landmarks & Includes a set of points representing the contour of the organs, as obtained by the HybridGNet. \\
Left Lung & Contains segmentation masks of the left lung in run-length encoding (RLE). \\
Right Lung & Contains segmentation masks of the right lung in RLE. \\
Heart & Contains segmentation masks of the heart in RLE. \\
Height & Height of the segmentation mask, necessary to decode RLE. \\
Width & Width of the segmentation mask, necessary to decode RLE. \\

\hline
\end{tabular}
\label{table:data_records_1}
\end{table}





\section*{Technical Validation}

The technical validation of the CheXmask dataset was conducted through a three-tiered approach: (1) evaluating the quality of HybridGNet segmentations by measuring the DSC on a gold-standard subset revised by an expert physician; (2) validating the relevance of the RCA-estimated DSC as performance metric; (3) evaluating the quality of the masks for the five source datasets separately, by exploring their suitability as a training GT set for the segmentation task.

\subsection*{1) Validation via physician annotations as gold-standard}

\begin{figure}[h!]
    \centering
    \includegraphics[width = 0.75\linewidth]{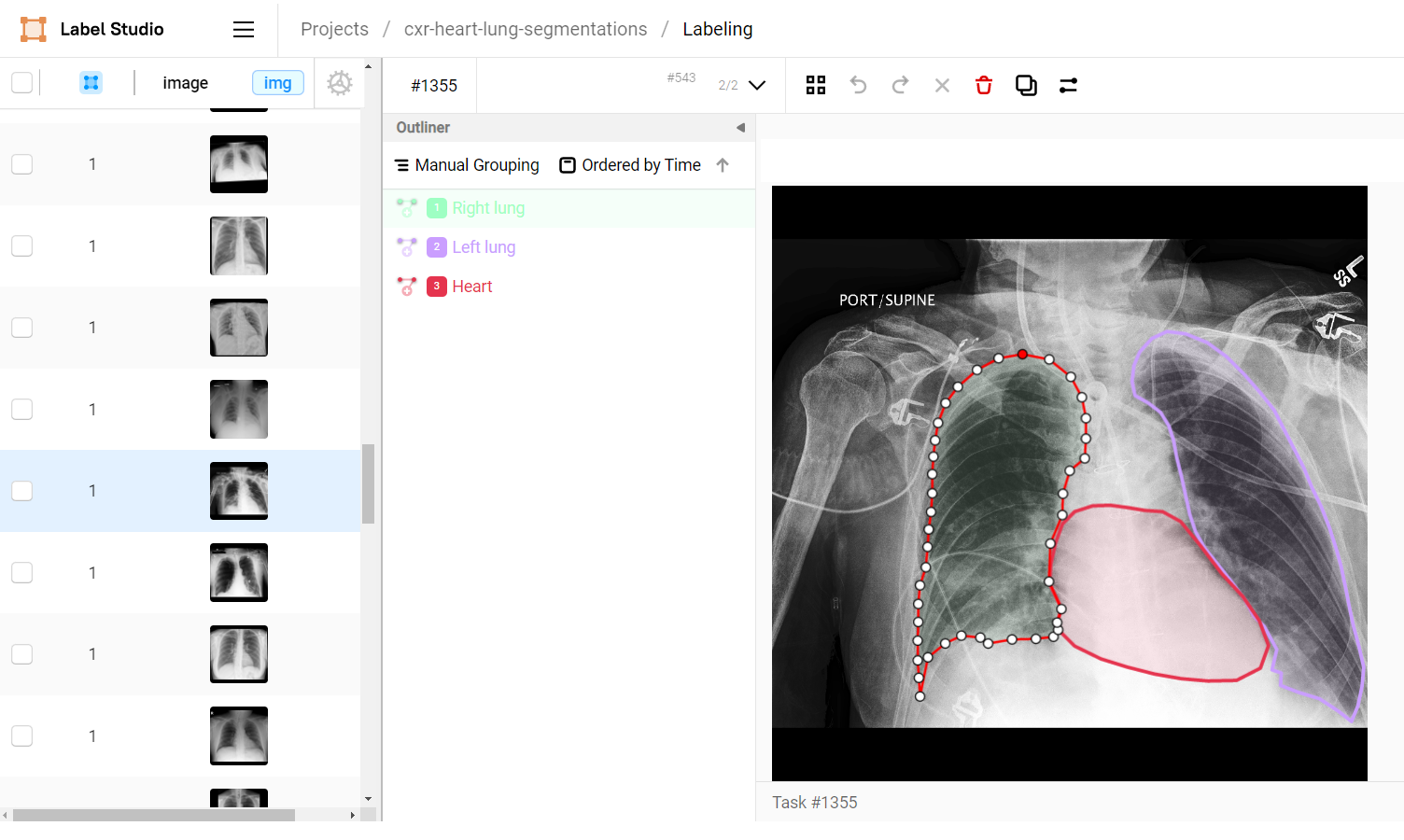}
    \caption{Label Studio setup for manual landmark-based segmentation. \textcolor{black}{Reproduction of x-ray images was allowed upon request to the original sources.}}
    \label{fig:labelstudio}
\end{figure}

We built a gold-standard set of masks (for images sampled from the five original datasets) \textcolor{black}{labeled by two experienced physicians, referred as P1 and P2,} to evaluate the segmentation quality of CheXmask lungs and heart masks. Both physicians, counting with more than five years of experience, performed a manual revision of the right lung, left lung and heart landmark-based segmentations. We used an open-source labeling platform called LabelStudio to do so. The landmarks predicted by HybridGNet for a subset of images were uploaded as predictions to LabelStudio, where the annotator corrected these predicted landmarks by moving them to their correct specific contour positions. This guarantees that the GT masks are based on the same collection of nodes than the evaluated masks, and reduces the time needed for annotation. Figure \ref{fig:labelstudio} shows the annotation interface. We include the source code to launch this interface in our repository, as an easy tool to perform further validations. We measured the DSC of HybridGNet's masks comparing with the manually-corrected landmark-annotations, and report statistics on the DSC distribution for the five public datasets. 

The sample size of the subset used as gold-standard evaluation was chosen based on the work by El Jurdi and Colliot \cite{eljurdi2023howprecise}, which suggests sampling 100 images for a 4\%-wide confidence interval in segmentation tasks. \textcolor{black}{We sampled 50 images per dataset, obtaining a final gold-standard set of 250 images. We combined random sampling, histogram-based sampling and a random selection of images which included radiological findings (disease or abnormalities) to ensure the presence of easy and challenging cases.} For each dataset, ten images were sampled based on a 10-bins histogram of the RCA-estimated DSC of all the dataset's images, taking one random image per bin. Another twenty images were then randomly selected from the whole dataset. \textcolor{black}{Finally, 20 additional images were taken from each dataset, ensuring the presence of pathology or findings such as edema, consolidation, pelural effusion or support devices, which usually result in more challenging segmentation scenarios.}

Table \ref{table_gt_validation} summarizes the results of physician validation for lung and heart segmentation across the five datasets on their 50-samples subsets. We report DSC, Hausdorff Distance (HD), and Hausdorff 95\% Distance (HD95) as evaluation metrics. HD and HD95 are distance error metrics, providing insights into the difference between the contours of the GT masks and the predicted masks. 

In terms of lung segmentation, the datasets generally exhibit high DSC values, ranging from 0.949 to 0.982. This indicates a strong overlap between the GT masks and the predicted lung masks. Among the datasets, VinDr-CXR exhibits the highest agreement for both physicians, while CheXpert has the lowest mean DSC. 

\textcolor{black}{For heart segmentation, the DSC ranges differ between the different annotators, in one case from 0.957 to 0.979 and, in the other, from 0.912 to 0.948. In both cases there is a strong overlap between the GT and predicted heart masks across all datasets (heart segmentation is a more difficult task).} The HD and HD95 values for heart segmentation are generally lower compared to lung segmentation, indicating a smaller distance between the contours.
\textcolor{black}{When comparing between the two expert physicians, the results vary in the same way as with the predictions. More difficult datasets such as CheXpert have the largest inter-reviewer differences, meanwhile VinDr-CXR has both better performance in each reviewer and the lowest inter-reviewer difference.}

Overall, the results indicate that the segmentation models perform well in capturing the lung and heart structures. VinDr-CXR consistently demonstrates the highest agreement for both lungs and heart, while CheXpert exhibits relatively lower agreement. These findings highlight the variations in segmentation performance across different datasets and provide valuable insights for further improving the accuracy of the segmentation algorithms.

\subsubsection*{Per-landmark error analysis}

In addition to the segmentation metrics, we also computed the mean squared error (MSE) per landmark across the entire set. This measure provided deeper insights into the positional accuracy of the landmarks predicted by HybridGNet. Interestingly, the results suggested that certain landmarks were more prone to prediction errors than others. Specifically, we observed higher MSE values for landmarks corresponding to the lower part of the lungs and the heart, indicating that these regions are more difficult for the network.

\begin{figure}
    \centering
    \includegraphics[width=\linewidth]{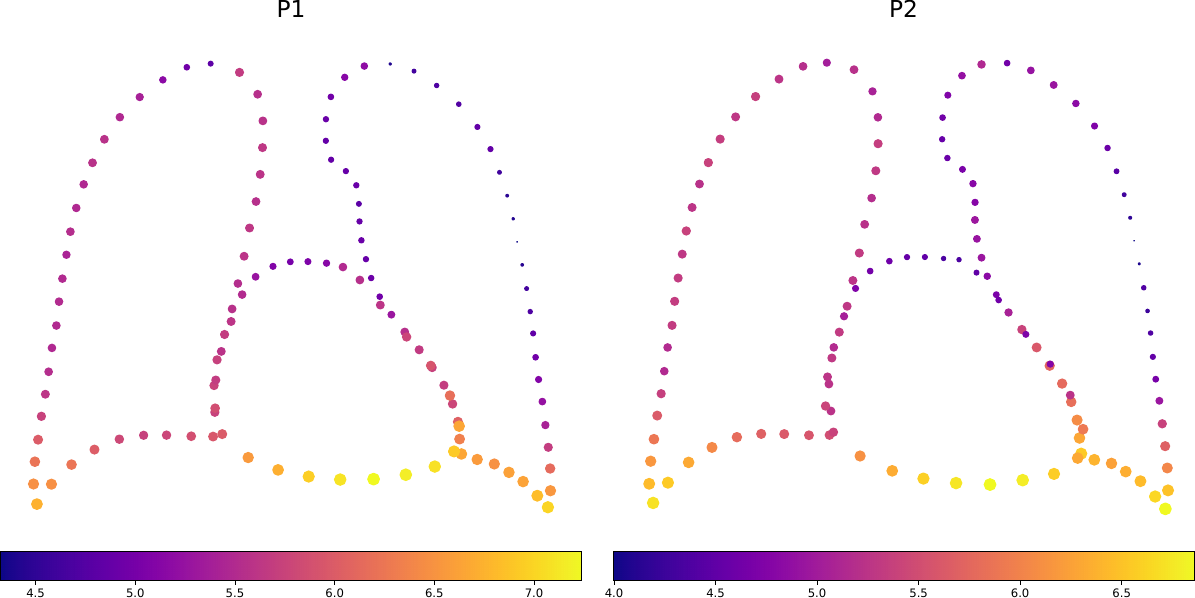}
    \caption{Illustration of the Mean Squared Error (MSE) per landmark across the entire gold-standard set \textcolor{black}{per expert reviewer}, depicted in logarithmic scale for improved visual clarity. The color intensity of each landmark represents the magnitude of the MSE. Moreover, the size of each landmark is proportional to its respective MSE, thus larger landmarks indicate a greater prediction error.}
    \label{fig:mse_per_landmark}
\end{figure}

Figure \ref{fig:mse_per_landmark} illustrates the MSE per landmark across the whole dataset \textcolor{black}{per expert reviewer}. Each landmark is color-coded based on its respective MSE value, using a logarithmic scale for better visualization. The size of each landmark also corresponds to the magnitude of the MSE, with larger landmarks indicating higher errors. As can be seen in this figure, the bottom part of the lungs and the heart landmarks tend to be larger, denoting higher prediction errors. This aligns with the observations made by \textcolor{black}{both expert physicians responsible for the manual annotations}, who noted that the boundaries of the heart and the lower section of the lungs tend to exhibit more blurriness compared to the upper part.

\begin{table}[ht]
\centering
\caption{Results of clinical validation. Comparison of predictions vs P1 and P2 ground truth annotations ($n=50$ per dataset)}
\begin{tabular}{lcccccc}
\toprule
 & \multicolumn{1}{c}{\textbf{DSC Lungs}} & \multicolumn{1}{c}{\textbf{HD Lungs}} & \multicolumn{1}{c}{\textbf{HD 95 Lungs}} & \multicolumn{1}{c}{\textbf{DSC Heart}} & \multicolumn{1}{c}{\textbf{HD Heart}} & \multicolumn{1}{c}{\textbf{HD 95 Heart}} \\
\midrule
\textbf{Pred vs P1} & & & & & \\
\textbf{CheXpert} & 0.949 (0.067) & 49.9 (49.7) & 30.0 (38.8) & 0.912 (0.083) & 40.8 (34.0) & 36.4 (32.6) \\
\textbf{ChestX-Ray8} & 0.966 (0.078) & 35.4 (53.7) & 18.6 (37.6) & 0.921 (0.111) & 34.8 (45.3) & 30.9 (43.1) \\
\textbf{MIMIC-CXR-JPG} & 0.962 (0.055) & 37.5 (40.4) & 21.1 (30.5) & 0.919 (0.093) & 36.1 (38.5) & 32.5 (36.6) \\
\textbf{Padchest} & 0.965 (0.067) & 36.7 (35.9) & 17.9 (23.0) & 0.948 (0.048) & 26.5 (23.1) & 22.7 (20.6) \\
\textbf{VinDr-CXR} & 0.982 (0.035) & 27.2 (33.0) & 10.2 (15.4) & 0.912 (0.098) & 39.8 (42.8) & 35.8 (41.3) \\
\midrule
\textbf{Pred vs P2} & & & & & \\
\textbf{CheXpert} & 0.958 (0.068) & 42.1 (49.4) & 24.3 (37.5) & 0.966 (0.070) & 21.5 (34.7) & 16.9 (28.5) \\
\textbf{ChestX-Ray8} & 0.969 (0.077) & 33.3 (51.6) & 16.6 (35.1) & 0.965 (0.105) & 18.2 (46.8) & 16.2 (44.0) \\
\textbf{MIMIC-CXR-JPG} & 0.962 (0.062) & 36.9 (42.6) & 18.3 (26.2) & 0.945 (0.083) & 30.1 (43.6) & 26.2 (39.6) \\
\textbf{Padchest} & 0.974 (0.070) & 27.2 (38.1) & 11.9 (22.6) & 0.979 (0.037) & 14.4 (26.4) & 11.7 (21.9) \\
\textbf{VinDr-CXR} & 0.982 (0.039) & 22.5 (35.3) & 8.0 (16.0) & 0.957 (0.102) & 21.1 (47.3) & 19.3 (44.5) \\
\midrule
\textbf{P1 vs P2} & & & & & \\
\textbf{CheXpert} & 0.962 (0.035) & 43.9 (55.9) & 23.0 (37.8) & 0.918 (0.064) & 40.4 (29.6) & 33.5 (25.0) \\
\textbf{ChestX-Ray8} & 0.976 (0.031) & 27.9 (25.5) & 12.2 (14.1) & 0.942 (0.064) & 24.5 (20.4) & 21.2 (19.0) \\
\textbf{MIMIC-CXR-JPG} & 0.966 (0.036) & 29.0 (21.8) & 15.8 (12.9) & 0.932 (0.059) & 31.5 (22.4) & 26.6 (19.9) \\
\textbf{Padchest} & 0.974 (0.027) & 30.8 (23.1) & 13.3 (12.5) & 0.951 (0.038) & 25.3 (19.8) & 20.8 (16.3) \\
\textbf{VinDr-CXR} & 0.983 (0.024) & 22.7 (18.5) & 8.1 (9.0) & 0.943 (0.037) & 25.3 (15.4) & 21.6 (14.5) \\
\bottomrule
\end{tabular}
\label{table_gt_validation}
\end{table}

\subsection*{2) Validation via RCA-estimated DSC} 

As previously mentioned, computing the RCA-estimated DSC requires training a rigid and deformable registration model based on deep learning. To this end, we employed 80\% of the ChestXray-Landmarks database for training. This model was used to propagate the predicted labels from the image atlas to the reference images (composed of the 5 most similar images to the atlas from the training set), and finally compute the RCA-estimated DSC (mean and max across the 5 values). 

\begin{figure}[h!]
    \centering
    \includegraphics[width = \linewidth]{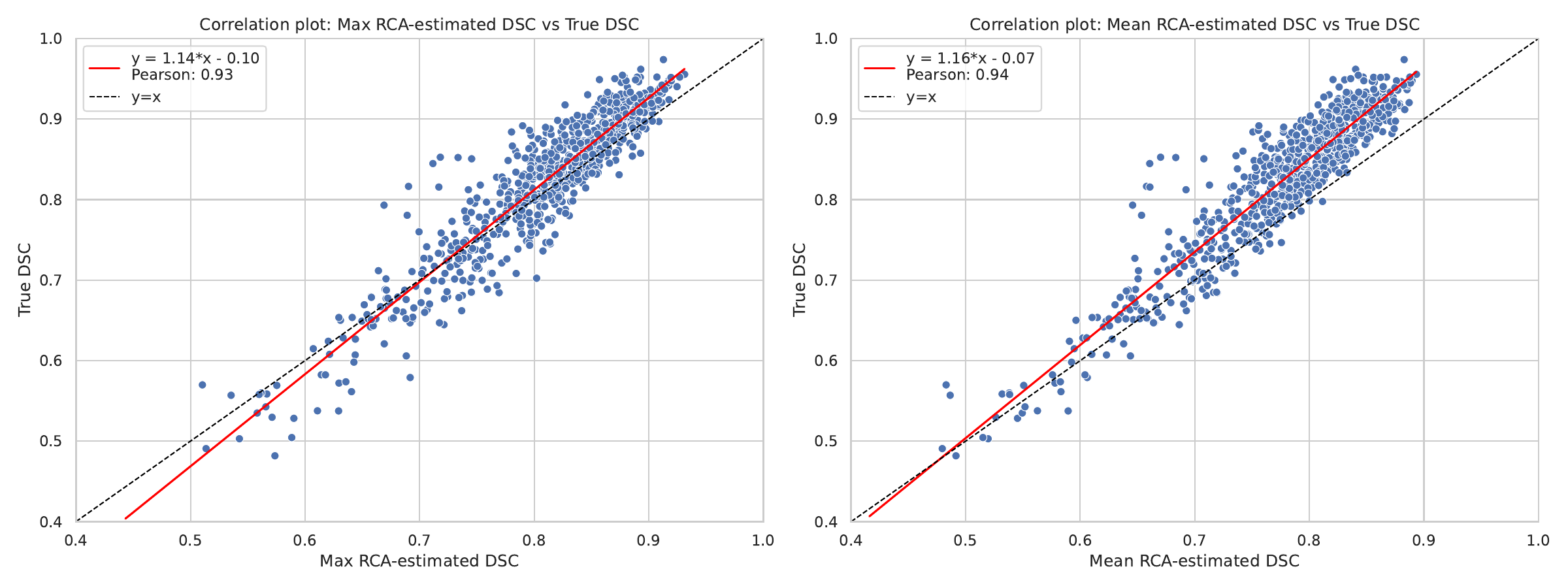}
    \caption{Correlation between the RCA values and the true DSC on the 20\% test-subset of the ChestXray-Landmarks database}
    \label{fig:correlation_plots}
\end{figure}

To validate that RCA-estimated DSC accurately approximates the real DSC, we proceeded to use simulated segmentation masks of various qualities for which we can compute the real coefficient. To this end, we generated a set of candidate segmentations of various quality (i.e. with real DSC ranging from low to high values), by training 12 UNet models on the same dataset but halting at various epochs (models trained for a few epochs produced lower quality segmentations while those trained until convergence produce better segmentations). Then, we assessed the correlation of RCA-estimated DSC on the remaining 20\%test-split of the ChestXray-Landmarks database. The Pearson Correlation scores for the Max and Mean RCA-estimated DSC across the reference set were 0.93 and 0.94 respectively, demonstrating a strong correlation (Figure \ref{fig:correlation_plots}). We found that the RCA-estimated value tends to slightly sub-estimate the true DSC, suggesting the performance measured with RCA follows a conservative approach (i.e., more pessimistic than the actual true performance). We selected the \emph{mean} RCA-estimated DSC as RCA estimation metric for the following experiments since it achieved better correlation.

\subsubsection*{Evaluation of segmentation quality through RCA}

\begin{figure}[h!]
    \centering
    \includegraphics[width = \linewidth]{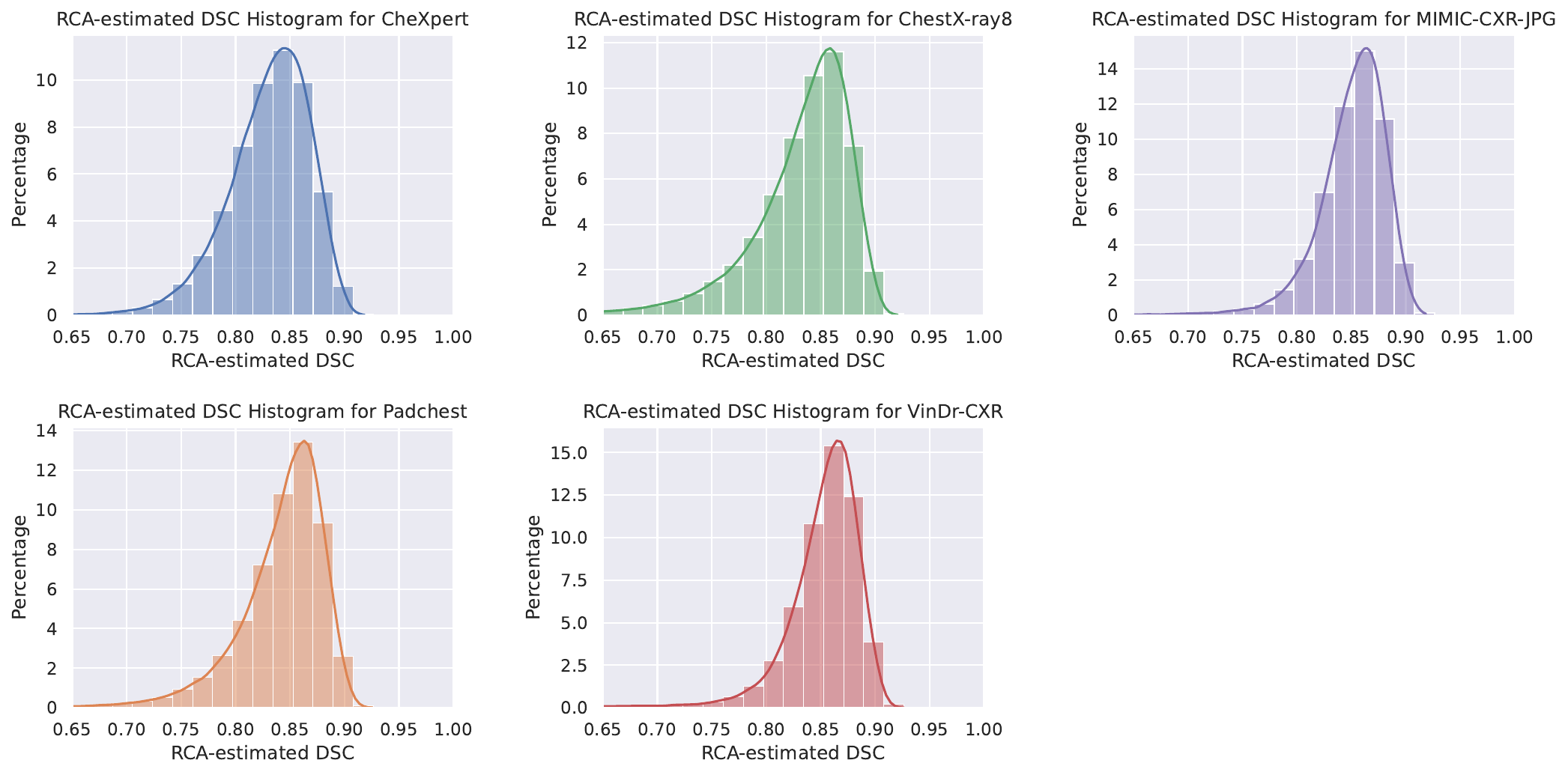}
    \caption{Histograms illustrating the distribution of segmentation quality across the five datasets. Each histogram represents the distribution of RCA (Reverse Classification Accuracy) estimations of DSC for a specific dataset, providing a visual representation of the segmentation performance. The histograms are truncated at 0.65 for visualization purposes. The full range of values is considered in the statistical analysis (Table \ref{table:rca_mean}).}
    \label{fig:histogram}
\end{figure}

As part of the technical validation on the quality of lungs and heart masks, we report a statistical description of the distribution of mean RCA-estimated DSC for all images in each dataset (Table \ref{table:rca_mean}). The full histograms are presented in Figure \ref{fig:histogram}. For all datasets, the mean RCA-estimated DSC is higher than 0.83. As explained above, the true DSC distribution probably has a larger mean, since the RCA-estimated DSC was found to under-estimate the performance.

\begin{table}[]
\caption{Mean image level RCA-estimated DSC: statistical analysis}
\begin{tabular}{ccccccccccc}
\hline
\textbf{Dataset name} &
  \textbf{Sample size ($n$)} &
  \textbf{Mean} &
  \textbf{Std} &
  \textbf{Min} &
  \textbf{1\%} &
  \textbf{5\%} &
  \textbf{25\%} &
  \textbf{50\%} &
  \textbf{75\%} &
  \textbf{Max} \\ \hline 
Chestx-ray8        & 112120 & 0.841 & 0.044 & 0.148 & 0.691 & 0.764 & 0.823 & 0.850 & 0.869 & 0.925 \\
CheXpert      & 187825 & 0.830 & 0.038 & 0.400 & 0.718 & 0.761 & 0.808 & 0.835 & 0.857 & 0.918 \\
MIMIC-CXR-JPG & 243334 & 0.831 & 0.051 & 0.149 & 0.650 & 0.738 & 0.811 & 0.842 & 0.864 & 0.921 \\
Padchest      & 96184  & 0.851 & 0.042 & 0.103 & 0.672 & 0.790 & 0.838 & 0.858 & 0.874 & 0.927 \\
VinDr-CXR     & 18000  & 0.850 & 0.033 & 0.469 & 0.738 & 0.794 & 0.835 & 0.855 & 0.872 & 0.919 \\
\hline 
\end{tabular}
\label{table:rca_mean}
\end{table}

\subsubsection*{Demographic bias analysis through RCA}

To explore potential biases \cite{larrazabal2020gender,ricci2022addressing}in the masks quality, we conducted a detailed analysis incorporating the metadata associated with the MIMIC-CXR-JPG, Padchest, ChestX-ray8, and CheXpert datasets. We took into account parameters such as disease findings, the sex and age of the patients, and the X-ray capture view (PA or AP).

\begin{figure}
\centering
\includegraphics[width = \linewidth]{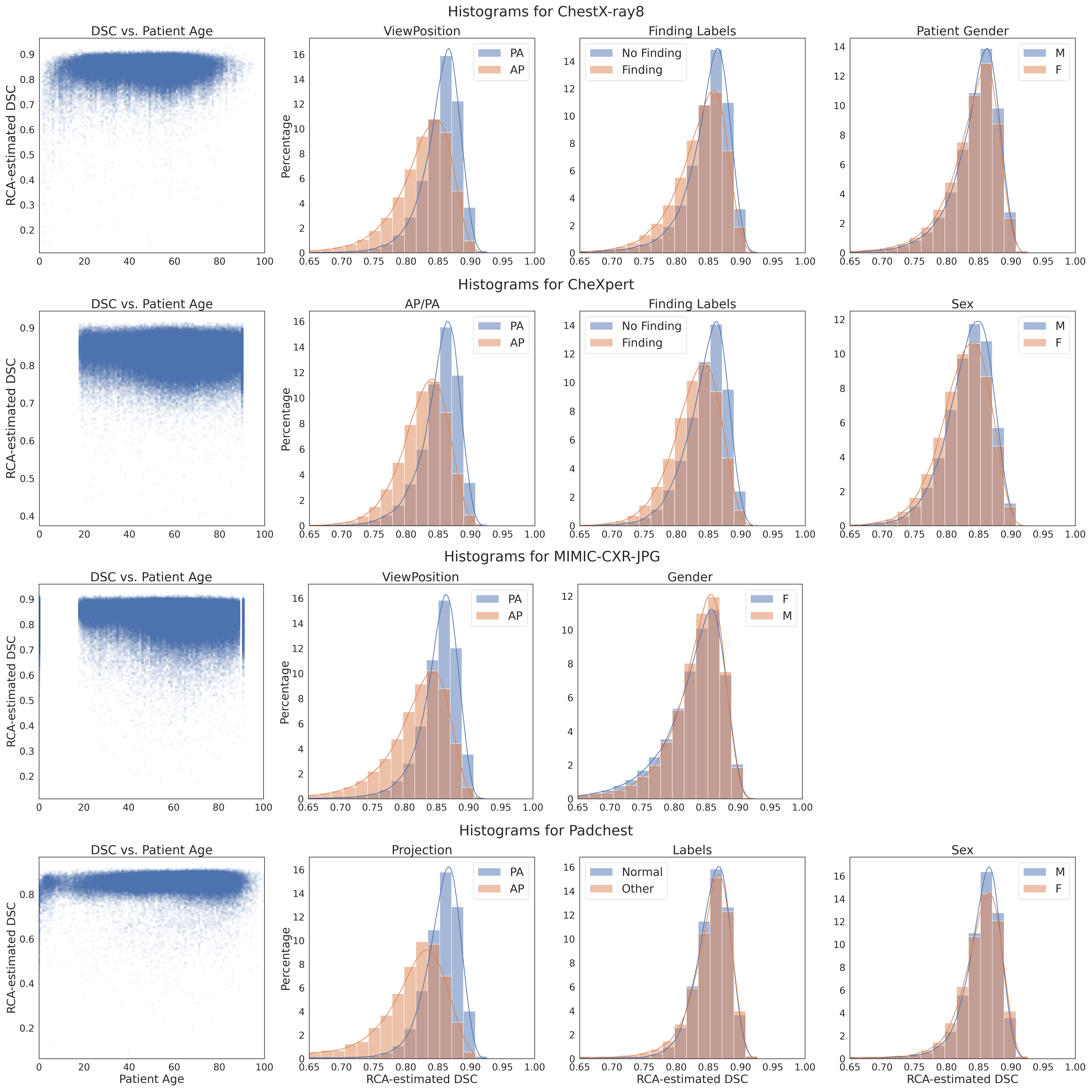}
\caption{Histograms depicting RCA-estimated DSC distribution across diverse metadata categories, namely imaging protocol, sex, and disease findings, supplemented by scatter plots correlating RCA-estimated DSC and age. These visual aids elucidate the relationship between segmentation quality and metadata categories, particularly highlighting differences associated with X-ray capture view (PA versus AP), age and the presence of disease findings.}
\label{fig:histograms-metadata}
\end{figure}

Figure \ref{fig:histograms-metadata} presents histograms that clearly demonstrate a superior RCA-estimated DSC for PA images compared to AP images across all investigated datasets. This is consistent with the fact that AP images tend to come from hospitalized patients, who are more difficult to position in standard views and usually include artifacts or cables that may act as confounders or occlude body parts, making anatomical segmentation more challenging. In contrast, discrepancies based on sex were less pronounced, with a slightly diminished segmentation quality observed in females. This is also consistent with the fact x-ray imaging of the upper thorax women’s breasts tend to occlude the imaged organs, resulting in poorer image contrast for the relevant anatomy \cite{ganz2021assessing} and making the segmentation task more challenging. Furthermore, the influence of disease findings on segmentation quality varied amongst datasets, with significant disparities evident in the ChestX-ray8 and CheXpert datasets.

Upon evaluating the relationship between patient age and RCA-estimated DSC in the CheXpert dataset, no substantial patterns emerged. It is important to note, however, that all subjects are above the age of 18. On the contrary, when analyzing results for the ChestX-ray8 dataset which includes patients under 18, we observed lower RCA-estimated DSC for these cases. This is consistent with results reported in a previous analysis of age patterns for the HybridGNet model \cite{gaggion2022}, where subjects below the age of 18 demonstrated subpar performance due to their absence in the training set. We hypothesize a similar trend may be affecting the sparse set of ChestX-ray8 X-rays taken for individuals under the age of 18.

These findings underscore the importance of desaggregated analysis with respect to image acquisition protocol, disease presence and demographic attributes when understanding segmentation quality. 

\subsubsection*{Out-of-distribution detection through RCA}

\begin{figure}
    \centering
    \includegraphics[width=\linewidth]{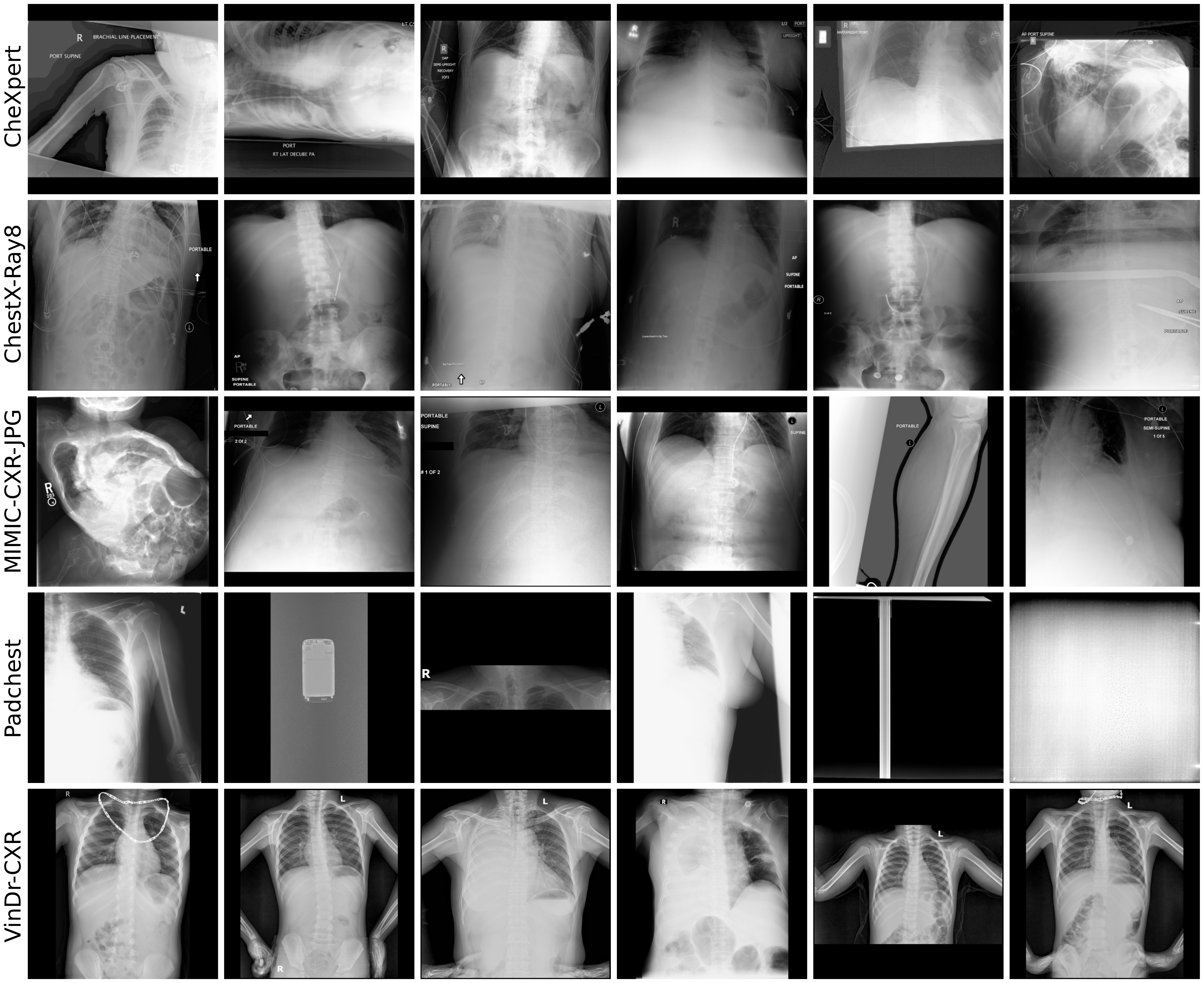}
    \caption{Illustration of a random subset of images with some of the lowest RCA values from each dataset. These images were identified as having low-quality segmentations based on their corresponding RCA scores. The selection showcases various types of images, including those with noise, mislabeled lateral CXR, and even non-medical objects like a mobile phone in the Padchest dataset. \textcolor{black}{Reproduction of these images was allowed upon request to the original sources.}}
    \label{fig:low_rca}
\end{figure}

In this section we focus on analyzing those segmentation masks which received lower RCA-estimated DSC values. Through empirical observation, we discovered that RCA-estimated DSC values lower than 0.7 tend to be indicative of out-of-distribution cases. To exemplify this, we randomly selected a subset of six images from each dataset with some of the lowest RCA values, which are showcased in Figure \ref{fig:low_rca}. By performing a manual and qualitative analysis of the lowest RCA-estimated DSC masks for each dataset, we found a wide variety of challenging scenarios, ranging from images with high noise levels or mislabeled lateral CXR, to an X-ray image of a mobile phone from the Padchest dataset. That is why we recommend users of the CheXmask dataset to only use segmentations whose RCA-estimated DSC is higher than 0.7.

Table \ref{table:rca_mean} can be useful to estimate the prevalence of such out-of-distribution images in each dataset. Notably, the VinDr-CXR dataset does not include any non-posteroanterior (PA) images nor many out-of-distribution images, as all images were manually annotated by physicians. However, the inclusion of zoomed-out images capturing both arms, which were absent in the HybridGNet training set, leads to lower-quality segmentations in those cases. In contrast, the larger datasets contain a higher number of cases mislabeled as PA or AP in their metadata. These mislabeled images could be easily identified and discarded by thresholding the RCA-estimated DSC, for example using the 1\% or 5\% quartile of a dataset as threshold. This approach can help to improve the overall quality and reliability of downstream applications developed using this datasets.

\subsection*{3) Validation via retraining segmentation models with CheXmask}

We performed a final quality control at the dataset-level to evaluate the quality of HybridGNet segmentations per dataset (not individually as we did with RCA-estimated DSC). The hypothesis is that if the segmentations are good, they can be used as training set for a segmentation model that will obtain high performance. Thus, we trained five new HybridGNet models, by using the set of predicted masks of each of the five public datasets as training GT. We then evaluated model performance on the complete ChestXray-Landmarks database. All models were trained using the same configurations and hyperparameters as the original HybridGNet model \cite{gaggionISBI2023}. As a comparison reference, we include the performance metrics reported on prior studies for the 20\% test split of the ChestXray-Landmarks database which can be seen as an upper bound in performance, as this model was trained with the training split of the ChestXray-Landmarks database.

Table \ref{rca_dataset} shows the results for lungs and heart segmentation. A slight decrease is observed in terms of DSC, decreasing from 0.967 to ~0.955 when using the segmented datasets as training set, in comparison to the reference values. This indicates that the overall quality of the segmentations is good enough to train models which perform well when compared with existing expert annotations.

\begin{table}[t!]
\centering
\caption{Dataset-level estimation of the quality of the segmentation via retraining the HybridGNet model from scratch.}
\resizebox{\linewidth}{!}{%
\begin{tabular}{ccccccc}
\hline
\multicolumn{1}{l}{}         & \multicolumn{3}{c}{\textbf{Lungs ($n = 911$)}}                   & \multicolumn{3}{c}{\textbf{Heart ($n = 383$)}}                   \\ \hline
\textbf{Training Dataset} &
  \multicolumn{1}{c}{\textbf{DSC}} &
  \multicolumn{1}{c}{\textbf{HD}} &
  \textbf{HD 95} &
  \multicolumn{1}{c}{\textbf{DSC}} &
  \multicolumn{1}{c}{\textbf{HD}} &
  \textbf{HD 95} \\ \hline
Chestx-ray8                         & \multicolumn{1}{c}{0.960 $\pm$ 0.019} & \multicolumn{1}{c}{36.2 $\pm$ 22.7} & 17.7 $\pm$ 10.3 & \multicolumn{1}{c}{0.927 $\pm$ 0.031} & \multicolumn{1}{c}{39.9 $\pm$ 16.5} & 35.2 $\pm$ 16.3 \\ 
CheXpert                       & \multicolumn{1}{c}{0.946 $\pm$ 0.028} & \multicolumn{1}{c}{42.0 $\pm$ 23.5} & 22.9 $\pm$ 12.1 & \multicolumn{1}{c}{0.888 $\pm$ 0.053} & \multicolumn{1}{c}{58.2 $\pm$ 27.0} & 54.0 $\pm$ 26.4 \\ 
Padchest                       & \multicolumn{1}{c}{0.956 $\pm$ 0.028} & \multicolumn{1}{c}{37.7 $\pm$ 23.7} & 19.0 $\pm$ 12.3 & \multicolumn{1}{c}{0.924 $\pm$ 0.037} & \multicolumn{1}{c}{40.9 $\pm$ 19.6} & 36.6 $\pm$ 19.2 \\ 
MIMIC-CXR & \multicolumn{1}{c}{0.956 $\pm$ 0.020} & \multicolumn{1}{c}{37.5 $\pm$ 22.8} & 19.2 $\pm$ 10.5 & \multicolumn{1}{c}{0.921 $\pm$ 0.031} & \multicolumn{1}{c}{44.2 $\pm$ 19.9} & 39.6 $\pm$ 19.6 \\ 
VinDr-CXR                     & \multicolumn{1}{c}{0.959 $\pm$ 0.023} & \multicolumn{1}{c}{35.4 $\pm$ 23.3} & 17.7 $\pm$ 11.7 & \multicolumn{1}{c}{0.925 $\pm$ 0.035} & \multicolumn{1}{c}{38.4 $\pm$ 16.8} & 34.0 $\pm$ 16.2 \\ \hline
\textbf{20\% test split} & \multicolumn{3}{c}{\textbf{Lungs ($n = 181$)}}                   & \multicolumn{3}{c}{\textbf{Heart ($n = 76$)}}                    \\ \hline
HybridGNet model trained on 80\% train split \cite{gaggionISBI2023} &
  \multicolumn{1}{c}{0.967 $\pm$ 0.017} &
  \multicolumn{1}{c}{32.5 $\pm$ 20.6} &
  15.6 $\pm$ 10.9 &
  \multicolumn{1}{c}{0.937 $\pm$ 0.029} &
  \multicolumn{1}{c}{34.6 $\pm$ 14.8} &
  29.4 $\pm$ 13.4 \\ \hline
\end{tabular}}

\label{rca_dataset}
\end{table}

\section*{Usage Notes}
The CheXmask dataset is intended to serve as a resource for researchers working in the field of medical imaging, particularly those interested in computational anatomy, shape understanding, CXR analysis and image segmentation. \textcolor{black}{The dataset opens avenues for various practical applications, such as deep learning model development and evaluation, training of generative models, clinical decision support systems, disease detection and diagnosis, and anomaly detection. Researchers can explore novel methodologies, such as the integration of masked autoencoders for self-supervised learning, wherein the anatomical segmentation masks serve as a rich source for unsupervised feature extraction, or in the implementation of attention and deep supervision techniques guided by masks, facilitating enhanced model focus on specific anatomical regions for improved localization and feature extraction. As an example of clinical application, the task of cardiothoracic ratio estimation, can be particularly enhanced by leveraging the precise anatomical information encoded in the segmentation masks, thereby contributing to more robust and interpretable cardiomegaly detection methodologies.}

For the sake of completeness, we release segmentation masks for all images included in our study. However, based on the qualitative and quantitative analyses presented in this paper (see section Technical Validation), our recommendation for downstream tasks is to only use segmentations whose RCA-estimated DSC is higher than 0.7, to avoid including out-of-distribution images as well as low quality masks. \textcolor{black}{In addition, more attention should be put on specific subgroups (like pediatric patients in ChestX-ray8 or images acquired in AP position) for which we observed lower RCA-estimated DSC values overall, as discussed in Section "Validation via RCA-estimated DSC". Please also note that, when using segmentation masks for downstream tasks (e.g. computing cardiothoracic ratio from heart and lung segmentation masks), additional validation considering ground-truth annotations for the particular task should be performed to ensure the quality of the segmentation masks is good enough for the task at hand (e.g. manual measurements of cardiothoracic ratio should be employed to validate the quality of the segmentation masks).}

A critical point to note is that this dataset does not include the original CXR images due to proprietary and privacy considerations. Hence, researchers intending to utilize this dataset are required to source the original images directly from the respective datasets cited in the manuscript. It is imperative that all conditions of usage and restrictions established by the providers of these datasets are fully respected and complied with.

The segmentation masks within this dataset are encoded using the Run-Length Encoding (RLE) technique, a standardized method for representing binary masks in a compact manner. In order to decode these masks back to their original binary format, researchers can use Python along with widely-used scientific computing libraries such as NumPy and Pandas. We provide the necessary scripts for encoding and decoding these masks to facilitate their use.

For researchers aiming to employ machine learning or deep learning techniques for analyzing this dataset, we recommend the use of well-established libraries. In particular, the HybridGNet model, used in this work, was developed in PyTorch. The source code for this model is made available alongside the dataset, allowing researchers to reproduce our results, conduct comparative studies, or further refine the methodology.

The segmentation masks are available in two formats. The first retains the original resolution as provided by the respective imaging source. The second consists of preprocessed masks with an uniform resolution across different imaging sources. We also provide the corresponding image preprocessing scripts which researchers can apply to the original images, enabling the use of all datasets at the same resolution. These scripts have been designed to replicate the preprocessing steps applied to the original datasets, thereby facilitating the integration of all these major CXR datasets.

\section*{Limitations of Study}

There are several limitations to consider in this study. Firstly, the validation of the subset of images was performed by a single physician, which may introduce potential biases or subjective interpretations. While efforts were made to ensure accuracy and consistency, further validation by multiple radiologists or experts could enhance the reliability and generalizability of the findings.

The selection process focused on including only CXR images in the posteroanterior (PA) or anteroposterior (AP) view, aiming to ensure consistency and homogeneity. However, due to potential inaccuracies or inconsistencies in the metadata (particulary in large-scale databases built from automatic analysis of electronic health records), there is a possibility that other types of images may have been included in the PA and AP views.
Despite the utilization of the Reverse Classification Accuracy (RCA) framework, which provides a reliable metric to detect low-quality segmentations, it is important to acknowledge that some images that do not meet the specific PA or AP criteria may have inadvertently leaked into the dataset. While RCA helps identify such instances, there remains a potential for the inclusion of non-PA or non-AP images that could impact the segmentation quality assessment.
Therefore, when working with these large datasets, it is crucial to exercise caution and acknowledge the possibility of mislabeled or misclassified images. Further investigations and refinements in the dataset selection process, including more robust metadata validation techniques, would be beneficial to enhance the reliability and accuracy of the segmentation results.


Finally, it is important to highlight that, since segmentation masks released in CheXmask are generated with HybridGNet, they tend to follow patterns of anatomical plausibility, even in presence of complex conditions like tuberculosis, which generate different type of organ oclussions. This is of particular importance for lung segmentation, where there are usually two ways in which masks can be annotated, either following the 'air' or 'anatomy' strategy \cite{anatomicalmasks,larrazabal2020post}.  While the 'air' strategy focus on segmenting the air cavity of the lung, excluding areas of increased opacity due to infection, the 'anatomy' strategy provide a comprehensive view which follows the expected anatomy, including these opaque areas. Segmentation masks generated by HybridGNet (i.e. those in CheXmask) are more similar to those generated via the 'anatomy' strategy. For a complete discussion about this, see Section V.5) and Figure 6 in Gaggion et. al \cite{gaggion2022}.

\section*{Code availability}

The code associated with this study is available in our Github repository: {\url{https://github.com/ngaggion/CheXmask-Database}. The repository encompasses Python 3 code for various components, including data preparation, data post-processing, technical validation, and the deep learning models. The data preparation section includes scripts for preprocessing the images. The data post-processing section provides scripts for converting the segmentation masks from run-length encoding to a binary mask format, examples of how to read the model and the necessary code to revert the pre-processing steps for each dataset. The technical validation section includes the code for the individual RCA analysis and the processing of the physician results. Additionally, the repository includes the code for the deep learning models used for image segmentation, including the HybridGNet model architecture, weights, training and inference scripts. The software prerequisites for running the code are outlined in the repository's README file.

\bibliography{bibtex}

\section*{Acknowledgements}

This work was supported by Argentina’s National Scientific and Technical Research Council (CONICET), which covered the salaries of E.F., D.M., L.M. and N.G. The authors gratefully acknowledge NVIDIA Corporation with the donation of the GPU computing used for this research and Agencia Nacional de Promoción de la Investigación, el Desarrollo Tecnológico y la Innovación (Agencia I+D+i PICT).

\section*{Author contributions statement}


E.F., N.G. and C.M. conceptualized the idea of this paper. N.G. performed the experiments. C.M. designed the LabelStudio interface and analyzed the results. L.M. implemented the deformable registration module for the RCA estimation. M.A. and J.M.S manually segmented the gold-standard set. D.H.M. and E.F. analyzed the results. All authors reviewed the manuscript.

\section*{Competing interests} 

The authors declare not conflict of interest.

\section*{Appendix: Statistical analysis for Max RCA-estimated DSC scores}

In the main manuscript we included statistical analysis for Mean RCS-estimated DSC scores since this indicator showed better correlation with real DSC. However, for the sake of completeness, here we include the same analysis for the alternative Max RCA-estimated DSC score. 

\begin{table}[h!]
\caption{Dice RCA (Max) statistical analysis}
\begin{tabular}{ccccccccccc} 
\hline
\textbf{Dataset name} &
  \textbf{Sample size ($n$)} &
  \textbf{Mean} &
  \textbf{Std} &
  \textbf{Min} &
  \textbf{1\%} &
  \textbf{5\%} &
  \textbf{25\%} &
  \textbf{50\%} &
  \textbf{75\%} &
  \textbf{Max} \\ \hline
Chestx-ray8        & 112120 & 0.882 & 0.044 & 0.164 & 0.732 & 0.806 & 0.864 & 0.890 & 0.909 & 0.967 \\
CheXpert      & 187825 & 0.870 & 0.038 & 0.415 & 0.757 & 0.801 & 0.849 & 0.875 & 0.896 & 0.961 \\
MIMIC-CXR-JPG & 243334 & 0.873 & 0.050 & 0.179 & 0.690 & 0.784 & 0.854 & 0.884 & 0.904 & 0.968 \\
Padchest      & 96184  & 0.893 & 0.042 & 0.128 & 0.714 & 0.831 & 0.882 & 0.901 & 0.917 & 0.970 \\
VinDr-CXR     & 18000  & 0.892 & 0.035 & 0.501 & 0.778 & 0.831 & 0.876 & 0.898 & 0.915 & 0.964 \\ \hline 
\end{tabular}
\label{table:rca_max}
\end{table}

\begin{figure}[h!]
    \centering
    \includegraphics[width = \linewidth]{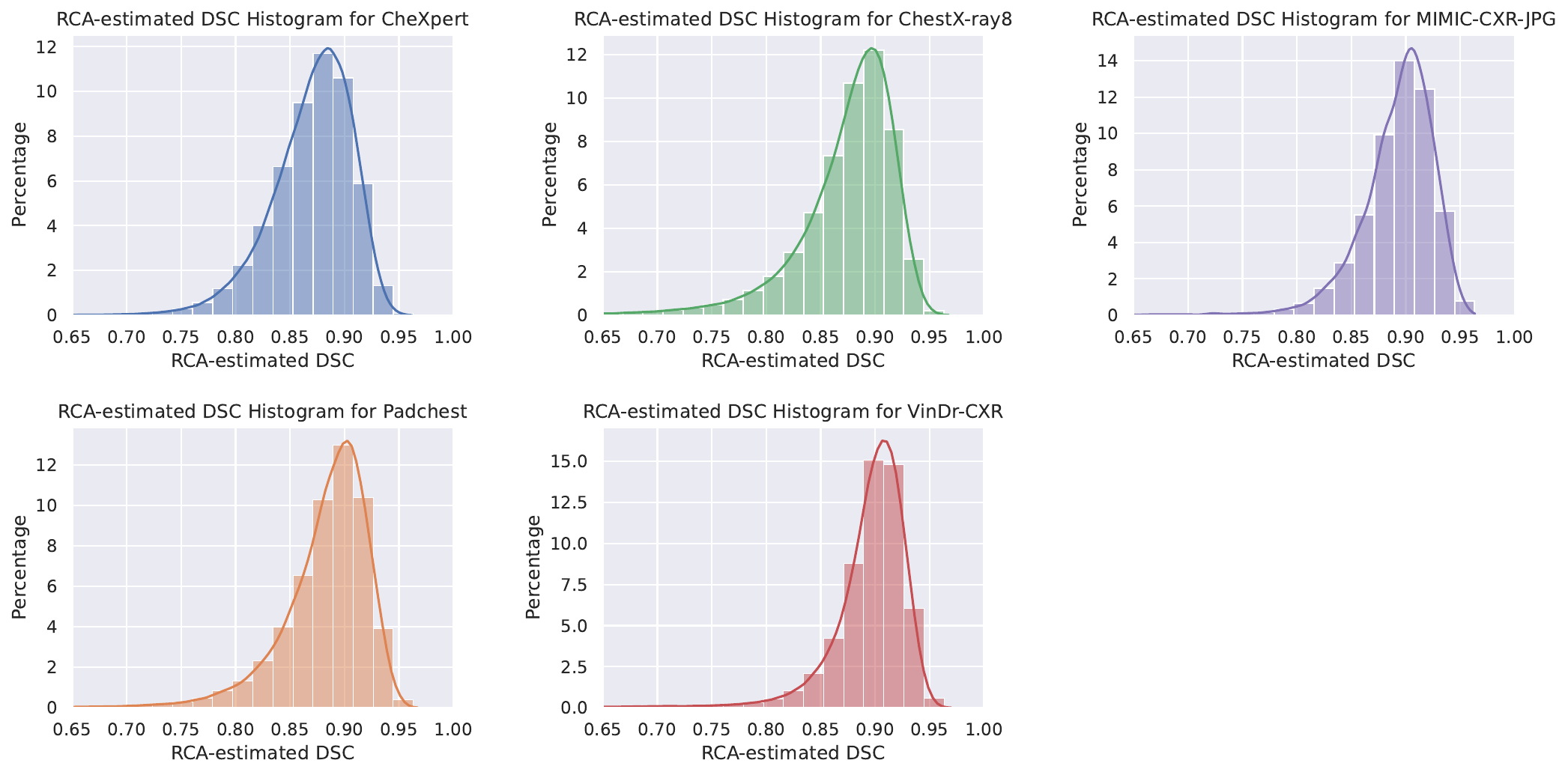}
    \caption{Histograms illustrating the distribution of segmentation quality across the five datasets. Each histogram represents the distribution of RCA (Reverse Classification Accuracy) estimations of DSC for a specific dataset using the Max value instead of the mean, providing a visual representation of the segmentation performance. The histograms are truncated at 0.65 for visualization purposes. The full range of values is considered in the statistical analysis (Table \ref{table:rca_max}).}
    \label{fig:histogram_max}
\end{figure}

\end{document}